# Information Sharing Among Countries: A Perspective from Country-Specific Websites in Global Brands

Amit Pariyar*, Yohei Murakami, Donghui Lin and Toru Ishida

*Department of Social Informatics*
*Kyoto University, Kyoto 606-8501, Japan*
*\*amit@ai.soc.i.kyoto-u.ac.jp; amitpariyar@gmail.com*

**Abstract.** Multiple official languages within a country along with languages common with other countries demand content consistency in both shared and unshared languages during information sharing. However, inconsistency due to conflict in content shared and content updates not propagated in languages between countries poses a problem. Towards addressing inconsistency, this research qualitatively studied traits for information sharing among countries inside global brands as depicted by content shared in their country-specific websites. First, inconsistency in content shared is illustrated among websites highlighting the problem in information sharing among countries. Second, content propagation among countries that vary in scales and coupling for specific content categories are revealed. Scales suggested that corporate and customer support related information tend to be shared globally and locally respectively while product related information is both locally and regionally suitable for sharing. Higher occurrences of propagation when sharing corporate related information also showed tendency for high coupling between websites suggesting the suitability for rigid consistency policy compared to other categories. This study also proposed a simplistic approach with pattern of sharing to enable consistent information sharing.

*Keywords*: Information sharing; global brands; country-specific websites; inconsistency; content propagation.

## 1. Introduction

Knowledge management is increasingly practiced among organisations to compete in the global market. The knowledge which is an invaluable asset to support such practice is often shown to have reference with information by the popular DIKW pyramid (Frické, 2009). Thus, past researchers have diverse opinion separating the terms "information" and "knowledge" but many agree on their relation. Rowley and Hartley (2008) stated knowledge as information having been processed, organised or structured in some way or else as being applied or put into action. Wallace (2007) stressed that contextual information is a prerequisite for knowledge. Therefore, information sharing is seen imperative to enrich knowledge, for example, technical know-how shared among employees internally or among offices situated in several countries (Almeida and Phene, 2004; Zhao and Luo, 2005; Adenfelt and Lagerström, 2006; Ryan *et al.*, 2010). Though digital platforms have encouraged sharing to maximum extent (Cormican and Dooley, 2007; Gupta *et al.*, 2009) inconsistency emerging from information that is conflicting, outdated and irrelevant are a possibility in environment where content is contributed in several languages globally. The importance of interaction and communication between those sharing and those receiving knowledge (Nohria and Ghoshal, 1994; Lagerström and Andersson, 2003; Steinheider and Al-Hawamdeh, 2004; He and Wei, 2009) also raises expectation for inconsistency during information sharing, a problem which is quite common in collaboration. This research focuses on inconsistency in information sharing which is problematic when it comes to enrich knowledge for supporting Knowledge management processes.

Past researches towards knowledge sharing in organisation have investigated primarily social factors, organisational commitment, technology and tool support for Knowledge management. Social exchange theory a commonly used theoretical base has elaborately explained trust, reciprocity and reputation as positive influences for knowledge sharing behaviours (Cabrera *et al.*, 2006; Liang *et al.*, 2008). Technology, wise the model of conversational knowledge management (Wagner, 2004) which is possible with web2.0 particularly Wiki technology seems supported commonly. Though the challenges to promote knowledge sharing are dealt in general, there is limited research on factors that could degrade knowledge. Given the growth of online communities and abundance of content in several languages, researches in information science on the other hand are inclined to address information disparities (Bey *et al.*, 2006; Adar *et al.*, 2009; Kumaran *et al.*, 2010; Monz *et al.*, 2011; Bronner *et al.*, 2012). The advanced techniques from NLP, self-supervised learning and machine translations are studied to support multilingual content synchronisation and change tracking between languages. The bulk of researches, however, are targeted at collaborative platforms such as Wikipedia where content is provided in diverse topics. In a business context, the scope of content is limited within an organisation and targeted at specific communities such as investors, customers, business partners therefore existing techniques cannot be directly applied to avoid inconsistency in information sharing, hence degrade knowledge. To gain a better understanding of information sharing in a business domain, this research targets content published in country-specific websites managed by global brands.

Associating countries and languages, this research also highlights a two-fold perspective on restrictions in content consistency during information sharing. First, the presence of multiple official languages within a country demands content consistency restricted only to its official languages. For example a country such as Switzerland with four official languages (Deutsch, English, French and Romansch) requires diplomatic document to be synchronised by legislation in all its official languages. Second, official languages common among countries either from geographic proximity or colonial history also put forth the necessity for content consistency in common languages. Table 1 gives a glimpse that besides "English" which is globally used "French" is an official language common in 29 countries worldwide and demanding consistent information sharing in as many as 21 countries in Africa. The occurrence of global events such as "Outbreak of Ebola Virus" is a perfect example to highlight how inconsistent content shared among Africa and Europe even in common language "French" is problematic during information sharing.

Table 1. Diversity in languages among geographical regions.

| Language   | World | Africa | Americas | Asia | Europe | Oceania |
|------------|-------|--------|----------|------|--------|---------|
| English    | 59    | 24     | 16       | 4    | 3      | 12      |
| French     | 29    | 21     | 2        | —    | 5      | 1       |
| Arabic     | 26    | 14     | —        | 12   | —      | —       |
| Spanish    | 21    | 1      | 19       | —    | 1      | 1       |
| Portuguese | 10    | 6      | 1        | 1    | 1      | —       |
| Russian    | 8     | —      | —        | 3    | 5      | —       |

Several issues during information sharing are apparent from such perspective. First issue refer to scales in sharing which means information is either restricted within specific country or shared in groups among several countries. Second is the publication and description of content either restricted to specific languages or shared among several languages. Avoiding such issues in information sharing degrades knowledge due to inconsistency as conflicting content or outdated content gets shared among countries. Previous technique for content synchronisation in Bronner *et al.* (2012) and Adar *et al.* (2009) has employed language processing to pinpoint overlapping or differential content between language but fails to take into account the scales and restriction in content being shared. Though multilingual content inconsistency due to technical limitation of machine translation is well known in Blenkinsopp and Pajouh (2010), Tanaka *et al.* (2011), Yamashita and Ishida (2006) and Yamashita *et al.* (2009), this study focuses on problems in sharing that fails to disseminate content consistently before even translation is needed.

To address inconsistency, this research qualitatively studies information sharing among countries inside global brands. For this purpose, country-specific websites that are published by global brands are used as a specific case of collaboration for information sharing among countries. First, content offered via country-specific websites is examined which reveals inconsistency in both shared and unshared languages among countries; an indication on problem during information sharing. Second, from examining content propagation among country-specific websites, traits in information sharing are revealed due to the presence of scales and coupling for specific content categories. To enhance knowledge, the rules for restrictions in the publication of content and their description in languages are then proposed for consistent information sharing.

This paper is organised as follows. Section 2 illustrates the problem in information sharing among country-specific websites with an example. The theoretical foundation including the hypothesis for this study is presented in Sec 3. Details concerning the methodology and the research findings are discussed in Secs. 4 and 5. The verification of hypothesis is followed by the proposal of mechanism for information sharing in Secs. 6 and 7. The contribution and limitations of this study are presented in Sec 8. Finally, we conclude in Sec 9.

## 2. Problem in Information Sharing

Global brands are multilingual by nature with offices situated in several geographic locations and customers with distinct cultural and language background. To cater to geographically dispersed customers, global brands have strategies for publishing content with country-specific websites targeted to specific country. With publication in more than 100 country-specific websites, global brands have huge volumes of content collaboratively generated in several languages (both shared and unshared languages) among countries. The occurrence of inconsistency in such collaboration is an indication for problem in information sharing. This section presents a motivating example to highlight inconsistency during information sharing.

Figure 1 illustrates inconsistency in the webpage "3M at a glance" published in country-specific websites managed by a global brand "3M". The collaboration among countries — United Kingdom (UK), Switzerland (CHE) and Canada (CA) — in their official language is also depicted with 'English' as a common language between UK and Canada while Canada and Switzerland both share a common language (French) and unshared language (Deutsch) respectively. Following problems are compiled in such collaboration.

### (a) Presence of outdated content

As illustrated in Fig. 1, the content updates for the year 2013 are not shared among the country-specific websites both in shared and unshared language. The content updates offered in country- specific website for Canada are not shared with UK even though they share a common language "English". Similarly, content updated in country-specific website for Canada is also not shared either in Deutsch (unshared language) and French (shared language) with Switzerland. The cases (i) and (ii) in Fig. 1 illustrate inconsistency from lack of propagation of content updates in the languages offered among the country-specific websites.

### (b) Presence of conflicting content

The case of content conflict is depicted between the languages (Deutsch and French) offered within a country-specific website for Switzerland from information about the "statistics for the number of employees". This case also depicts inconsistency occurring only for languages used within a country that are targeted for domestic customers (shown as (iii) in Fig. 1). In addition to this, the content in shared language (French) offered at country-specific websites for Canada and Switzerland also shows possibility for conflicts occurring between countries from the absence of propagation of content updates.

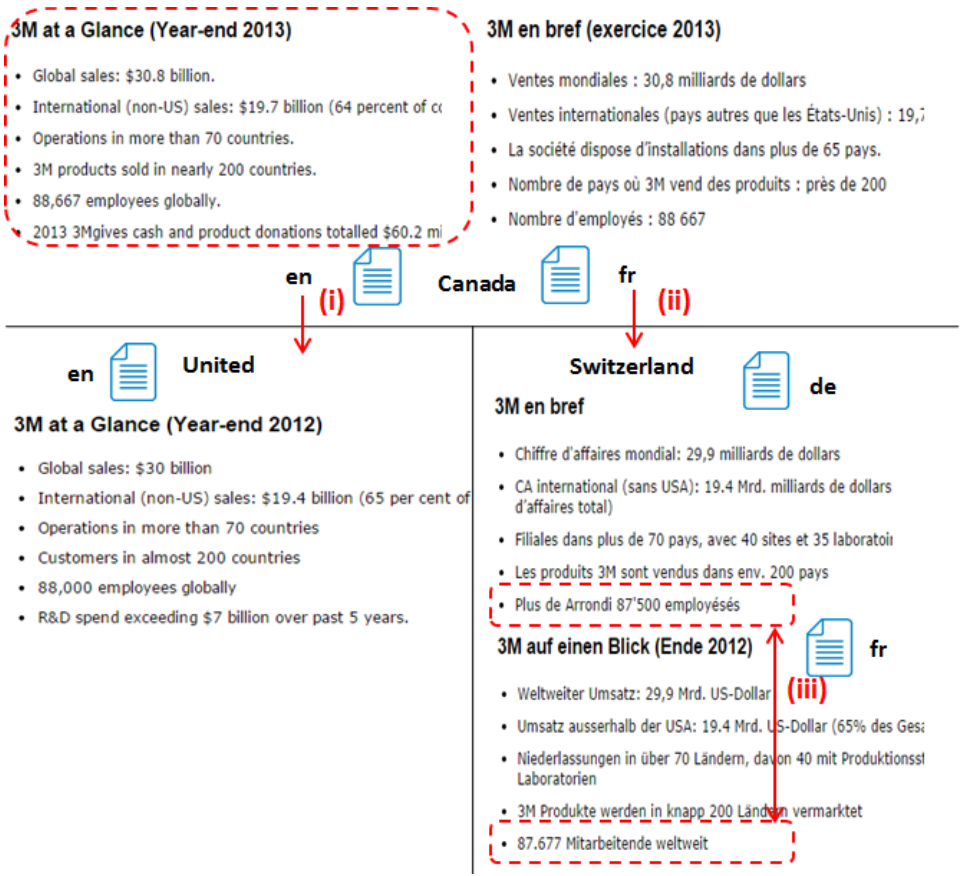

Fig. 1. Example of inconsistency in content shared among country-specific websites. (i) and (ii) Updates not propagated in shared and unshared language, (iii) content conflicts.

Inconsistency from lack of propagation of content updates or the presence of outdated content and content conflicts among country-specific websites are indications for problems in information sharing among countries. Without consistent information sharing, knowledge is degraded which is detrimental to knowledge management processes. In the presence of problems during information sharing, the following question arises.

**RQ 1.** Do online platforms such as country-specific websites have traits embedded for sharing content belonging to specific categories?

The goal from this research is to expand our understanding of information sharing among countries and examine possible traits embedded in content shared via country-specific websites. The purpose is towards enriching knowledge with information sharing that is consistent across shared and unshared languages. Finally, this study also aims to propose a mechanism that enables consistent information sharing. The next section presents the theoretical foundation and hypothesis for this study.

## 3. Theoretical Background and Hypothesis

This section presents insight on the challenges for consistent information sharing due to language diversity and channels of sharing in an organisation. The perspectives from globalisation studies are also discussed to develop hypothesis for investigating the research question posed in this study.

### 3.1. *Managing language diversity*

Managing geographically and culturally diverse network of organisational units such as foreign overseas offices is complicated with language barrier (Charles and Marschan-Piekkari, 2002; Welch *et al.*, 2001). Severe communication problems were reported in global manufacturing of Fiat motorcar as logistic personnel along the supply chain struggled to work in a mix of Spanish, Portuguese, English, Italian, Polish and Russian languages (Feely and Harzing, 2003). Such problems show the complexity added to collaboration due to diversity, penetration and sophistication of the languages used within an organisation. To avoid such problems corporate language is often used for business purposes but it limits the reach to few languages and communities. By using specific language for business purposes, information sharing cannot scale to wide networks (Marschan-Piekkari *et al.*, 1999; Fredriksson *et al.*, 2006) of customers, business partners and investors with distinct language backgrounds. The resistance in the communication among overseas offices due to adoption of corporate language alien to the overseas employees restricts information sharing. Other issues while managing language diversity are the technical complexities involving machine translations with asymmetries, intransitivity, untranslatable words as barriers in MT mediated collaboration in any multinational project (Yamashita and Ishida, 2006; Inaba *et al.*, 2007; Yamashita *et al.*, 2009; Blenkinsopp and Pajouh, 2010; Tanaka *et al.*, 2011).

### 3.2. *Channels for sharing*

While language diversity is a huge barrier for collaboration it is essential to discuss the effectiveness of channels used for information sharing. In an intra-organisational setting where employees work at the same location, face to face interactions are shown to be more effective for sharing (Markus, 2001). However, the reliance on collaborative tools for sharing is increased with the increasing geographic distances. Just as with face to face interaction, the temporal and freshness of information are also important to achieve with collaborative tools. Therefore, prior to dealing with languages, disseminating up-to-date information where necessary is of utmost concern when implementing channels for sharing. Language issues can be dealt with technologies such as machine translation as a second step only after the dissemination of up-to-date information is guaranteed with the channel used for sharing.

### 3.3. *Sharing via country-specific websites*

The web has primarily served as the favoured channel by organisation to share information beyond socio-geographic limitations. Be it for interacting with the customers, business partners, suppliers, investors and so on, the websites for Business to Business (B2B) as well as for Business to Customer (B2C) are suited for global presence (Yunker, 2002; Singh and Pereira, 2005; Singh *et al.*, 2005). Global websites appear as strategic response of organisations where globalisation has a deep rooted influence in the design and cultural adaptation to target audiences (Lionbridge, 2009). In the web globalisation report card (Yunker, 2014), the distinctive feature of the best global websites such as increasing support for the languages, global reach with a local touch, accessibility of content from organised global gateways and so on are illustrated as pre-requisites for global presence. However, the presence of country-specific websites (Daryanto *et al.*, 2013) targeting individual countries in their official languages also stand out bringing forward the concept of information sharing occurring among the countries.

#### 3.3.1. *Cultural influences in sharing*

As diverging perspectives on cultural influences persist in the design of websites towards meeting the locale specific need of the countries, it is interesting to relate cultural influences in sharing information among countries. Where the view on cultural homogeneity among countries favours standardisation of product and services across the globe (Hall, 1997; Main, 2001) the standardisation in information (content offered in webpage from websites) sharing between countries is yet to be investigated. Similarly, where majority of researches have lenience towards Hofstede typology of cultural differences existing among the countries (Hofstede, 1984; Kale, 1991), information (content offered in webpage from websites) sharing is also expected to be influenced from such differences.

#### 3.3.2. *Embedded traits in sharing*

Confounding to homogenising effect, previous studies (Tixier, 2005; Singh *et al.*, 2009) have presented website categories: from standardised to highly localised websites to depict differences in web globalisation efforts placed in global organisations. While standardised websites are an effort towards global design template; localisation is an effort towards design adapted for specific localities. Similar website categories in Maynard and Tian (2004) also depict the levels of cultural adaptation with global appeal and local touch as glocal websites.

Bringing this notion into country-specific websites that are managed by global brands, the sharing of content among the websites presumably either depicts standardisation with same content offered for both domestic and international users or localisation with content localised for international users and not shared among the country-specific websites. However, the presence of content in varying proportion among the websites indicates on sharing of content restricted to specific websites. As several categories of content are published in websites, the restriction in their sharing is presumed to occur due to varied suitability among countries. Previous researches (Huizingh, 2000; Robbins and Stylianou, 2003) have noted the differences in the content and design features among cultural groups. The clue for differences in the suitability for sharing due to content categories is provided from the propagation of content among the country-specific websites. Therefore, to shed light on differences in suitability for sharing, we set the following hypothesis to examine propagation.

**Hypothesis 1.** *Propagation among county-specific websites is constrained by content categories: Corporate Information, Product Information, and Customer Support Information.*

The goal from the stated hypothesis is to uncover embedded traits in information sharing inside global organisations as depicted by content shared among their country-specific websites. The contribution will eventually be towards proposing solution for consistent information sharing that enriches knowledge, a valuable asset in organisation. Next, the methodology is discussed.

### 4. Research Methodology

This section details the methodology for this qualitative study on information sharing. Figure 2 gives an outline of examining sharing among countries from sampling webpages to comparing webpages among country-specific websites in global brands.

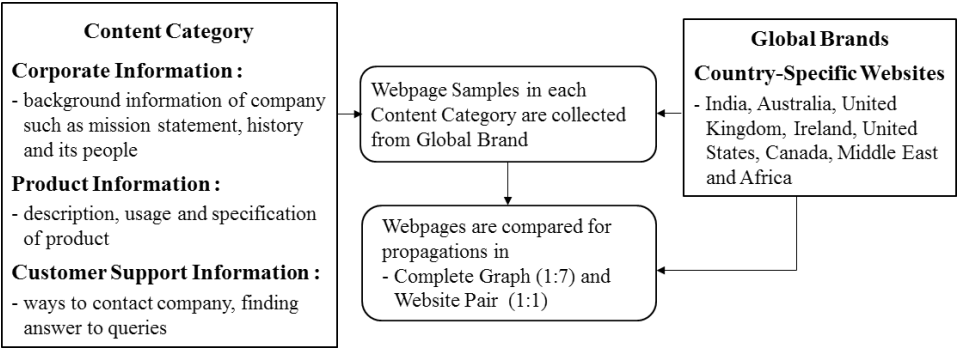

Fig. 2. Outline for examining content shared among countries.

### 4.1. *Data source*

Websites from 10 global brands that are ranked highly in the web globalisation report card (Yunker, 2014) are selected for this study. Each of the chosen global brand offers worldwide product and services with webpages published in more than 40 country-specific websites and more than 20 languages (Table 2). For this study, webpages offered in shared language (English) among the country-specific websites are selected. A sample of eight country-specific websites in each global brand representing countries — India (IN), Australia (AU), United Kingdom (UK), Ireland (IE), United States (US), Canada (CA), Middle East (ME) and South Africa (ZA) from several geographic regions: Asia Pacific, North America, Europe and Middle East-Africa — are selected. A total of 80 country-specific websites are collected as the source for webpages to be used for comparison (Table 3).

From eight country-specific websites, there are 28 possible websites pairs representing content sharing in country pairs for each global brand. For example, sharing among India and remaining countries occur in website pairs as: (IN, AU), (IN, UK), (IN, US) and so on.

### 4.2. *Content category*

Previous researches (Huizingh, 2000; Robbins and Stylianou, 2003) presented content features with categories that provide general company information, financial information, support and employment information to the customer and so on. Such

Table 2. Statistics on country-specific websites of global brands.

| Global brand | Industry | Country-specific websites | Sample websites* |
|---|---|---|---|
| Nivea | Skin and body care | 70 | 8 |
| 3M | Conglomerate | 100 | 8 |
| Starbucks | Coffee shop | 41 | 8 |
| Acer | Computer | 60 | 8 |
| Samsung | Conglomerate | 143 | 8 |
| KPMG | Professional services | 143 | 8 |
| HP | Computer hardware electronics | 88 | 8 |
| Nestle | Food processing | 75 | 8 |
| Avon | Personal care | 74 | 8 |
| John Deere | Heavy equipment | 63 | 8 |

*Note*: *Country-specific websites selected from several geographical regions.

Table 3. Statistics on websites and webpages.

| Brands | Website in each brand | | Content category | Webpage samples in each category | Webpage in each brand | Total | |
|---|---|---|---|---|---|---|---|
| | Individual | Pair | | | | Individual websites | Webpage |
| 10 | 8 | 28 | 3 | 16 | 3 * 16 = 48 | 10 * 8 = 80 | 10 * 48 = 480 |

features are associated with the design and cultural adaptation in the corporate websites. This study also uses content categories in sampling webpages from each global brand which are (a) Corporate Information: in sampling webpages that provide background information of a company such as mission statement, history and its people, (b) Product Information: in sampling webpages on description, usage, and specification of product and (c) Customer Support Information: in sampling webpages on ways to contact company or find answer to queries.

### 4.3. *Webpage samples*

Webpages from each global brand are manually analysed to label them to specific content categories: Corporate Information, Product Information and Customer Support Information. As shown in Table 3, 48 webpage samples are collected in each global brand making a total of 480 webpage samples used for this study.

### 4.4. *Comparison of webpage*

From webpage samples, the content in webpages is qualitatively compared to determine whether propagation occurs or does not occur among the websites. A paragraph of text in a webpage of a selected Website is used as a threshold to check for its presence among the remaining websites. Propagation is said to occur among the websites upon the presence of exactly the same paragraph or comparable paragraph in their webpages. Comparable paragraphs are paraphrased text that provide same information in the webpages of corresponding websites. Similarly, no propagation among websites is assigned when content is not same in the webpages of the corresponding website or when webpages do not exist. Propagation among country-specific websites is examined in a complete graph and in website pairs which are explained below.

#### 4.4.1. *Complete graph*

In the absence of publicly accessible information on the specific website where the content originates in a webpage, each country-specific website is considered as a potential source for publishing content in a webpage and sharing with the remaining websites. Figure 3(a) is an example depicting country-specific website for India chosen as a potential source for publishing content in a specific webpage. The presence of same or comparable content in corresponding webpages residing in the remaining country-specific websites such as Australia, United States, Canada and so on illustrates the propagation from India to remaining countries i.e. propagation occurs in (1: 7) country-specific websites.

For this study, propagation in (1: 7) country-specific websites is analyzed from 8 potential sources. The comparison of webpages performed from all potential source websites represents a complete graph as shown in Fig. 3(b). A total of 480 webpages are qualitatively compared for their propagation in (1:7) country-specific websites (Table 4). The purpose of studying propagation in (1 : 7) country-specific websites is to examine the presence of scales in sharing content for specific content categories.

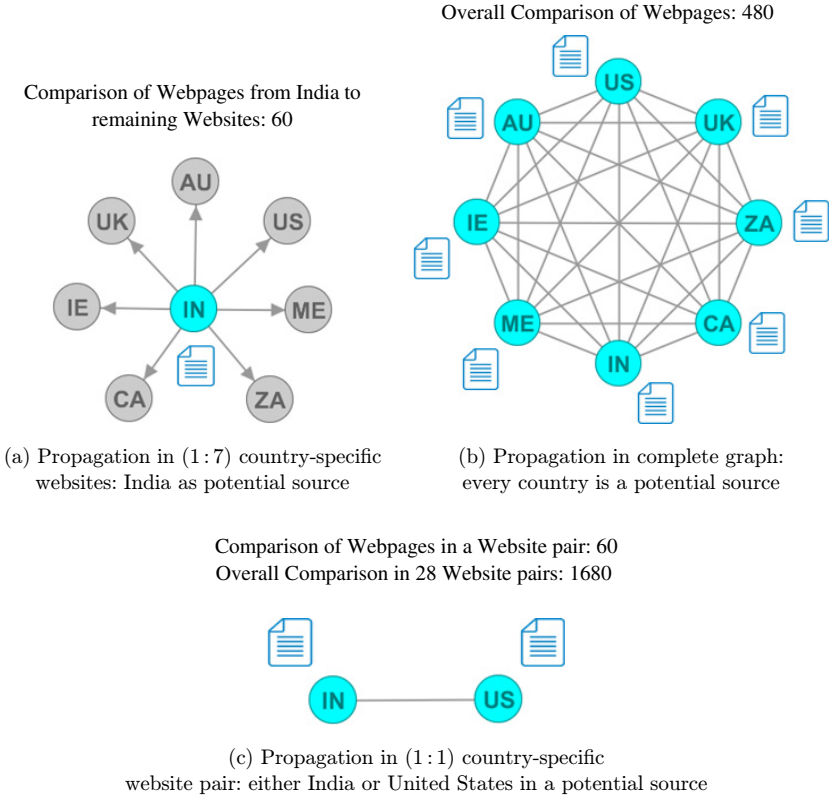

Fig. 3. Comparison of webpages for propagation among country-specific websites. *Cyan colored node is the source.

### 4.4.2. *Website pair*

Propagation in (1:1) country-specific website pair is also examined by comparing the content in the webpages of the corresponding websites (Fig. 3(c)). As shown in

Table 4. Statistics on comparison among country-specific websites.

| Complete graph | | | | 1 : 1 Website pair | | | |
|---|---|---|---|---|---|---|---|
| Each category | | All category | | Each category | | All category | |
| Each 1: 7 | Complete | Each 1: 7 | Complete | Each pair | All pair | Each pair | All pair |
| 20 | $20*8$ $=160$ | $3*1*20$ $=60$ | $3*8*20$ $=480$ | 20 | $28*20=560$ | $3*1*20=60$ | $3*28*20$ $=1680$ |

Table 4, for each website pair there are 60 comparisons of webpages making a total of 1680 comparison for all 28 website pairs. The purpose of studying propagation in (1:1) country-specific website pair is to examine coupling between websites in sharing content for specific content categories. The next section details the findings.

## 5. Research Finding

The results of comparing webpages among country-specific websites are detailed here. Interesting results for scales and coupling in sharing for specific content categories are compiled from examining their propagation.

### 5.1. *Propagation in complete graph*

Tables 5 and 6 present quantitatively the result of comparing webpages for propagation in (1:7) country-specific websites over a complete graph. The suitability of content globally, regionally and locally is represented with three cases: (a) propagation to all country-specific websites (b) propagation to some country-specific website and (c) no propagation. Scales in sharing are determined from such cases for specific content categories. Differences in scales while sharing content for specific content categories are revealed from comparing their propagation.

As illustrated in Table 6, out of 160 comparisons of webpages in "Corporate Information", 50% of cases are identified in which propagation occurs among all country-specific websites while 32% of cases in propagation occurs in some websites and in 18% of cases no propagation occurs among the websites. With more than 80% cases in which propagation occurs from at least a single country-specific website, the suitability of content in "Corporate Information" is not limited within a single country. The result strongly suggests for the suitability of corporate related information globally in all country-specific websites.

Comparing webpages in "Product Information" revealed suitability mostly either for some countries or limited to a specific country. Only 15% cases in which

Table 5. Propagation for content categories in complete graph.

| Propagation in 1:7 website* | Corporate information | | | Product information | | | Customer support information | | |
|---|---|---|---|---|---|---|---|---|---|
| | All | Some | None | All | Some | None | All | Some | None |
| IN | 10 | 7 | 3 | 3 | 7 | 10 | 4 | 3 | 13 |
| AU | 10 | 7 | 3 | 3 | 6 | 11 | 4 | 3 | 13 |
| UK | 10 | 9 | 1 | 3 | 10 | 7 | 4 | 4 | 12 |
| IE | 10 | 8 | 2 | 3 | 9 | 8 | 4 | 4 | 12 |
| US | 10 | 3 | 7 | 3 | 4 | 13 | 4 | 1 | 15 |
| CA | 10 | 6 | 4 | 3 | 5 | 12 | 4 | 1 | 15 |
| ME | 10 | 5 | 5 | 3 | 8 | 9 | 4 | 3 | 13 |
| ZA | 10 | 7 | 3 | 3 | 8 | 9 | 4 | 4 | 12 |
| Total | 80 | 52 | 28 | 24 | 57 | 79 | 32 | 23 | 105 |

*Note*:* Combination of all 1:7 propagation equals a complete graph.

Table 6. Summary on propagation cases for content categories.

| Content category | Propagation in country-specific websites | | | | | | Comparison of webpage |
|---|---|---|---|---|---|---|---|
| | All | % | Some | % | None | % | |
| (a) Corporate information | 80 | 50 | 52 | 32 | 28 | 18 | 160 |
| (b) Product information | 24 | 15 | 57 | 36 | 79 | 49 | 160 |
| (c) Customer support information | 32 | 20 | 23 | 14 | 105 | 66 | 160 |
| Total | 136 | 28 | 132 | 28 | 212 | 44 | 480 |

propagation occurs in all websites are identified which strongly suggests that product related information is not globally suitable. However, 36% cases of propagation to some websites and 49% cases of no propagation are comparable to infer the suitability of product related information both regionally and locally among countries. Contrary to this, comparing webpages in "Customer Support Information" strongly suggested the suitability of support related information locally within a country with 66% cases of no propagation among countries.

Figure 4(a) presents a comparative view on propagation for the content categories in a complete graph. Customer Support Information tends to be favoured over Product Information for propagation among all country-specific websites. Both Corporate Information and Product Information tend to be comparatively favoured for sharing in some countries while Product Information also shows preference for

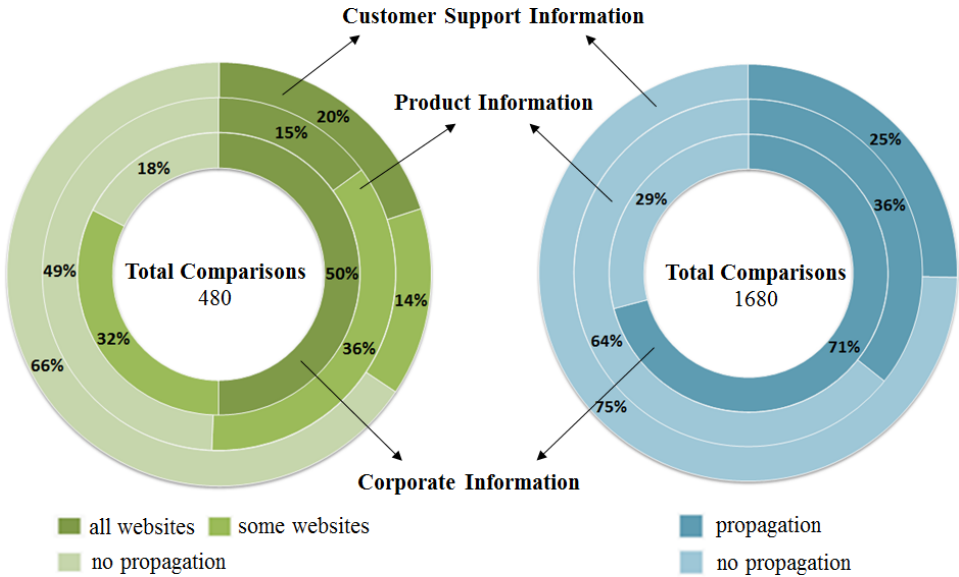

(a) Propagation in complete graph     (b) Propagation in (1:1) website pair

Fig. 4. Propagation for content categories in global brands.

sharing only within a country. From examining propagation in (1:7) websites among all country-specific websites, some websites or within a website; the scales in sharing for specific content categories are compiled in this study.

### 5.1.1. *Scales in sharing*

Propagation of "Corporate Information" at a global scale suggests the sharing of corporate related information among global communities. In such a case, the contribution of information is permitted to occur from the participation of local communities while dissemination of up-to-date information is to occur globally for consistent information sharing among global communities.

Propagation of "Product Information" both at regional and local scale suggest dissemination of up-to-date information either restricted among several countries within and across regions or limited to specific country when describing product specifications, usage and so on.

Propagation of "Customer Support Information" at a local scale suggests dissemination of information restricted to local communities. As it is logical that suitability of content is limited to specific locale where it is produced, local scale also suggests for synchronisation of content updates to occur within a country, for example, content synchronised in official languages (English and French) within a country for Canada.

Scales in sharing globally, regionally and locally are also useful to enable content consistency in information sharing. The restriction in publication of content and their description in specific languages associated with scales are useful in disseminating consistent information belonging to specific content categories. Next, results from comparing webpages in website pairs are compiled.

## 5.2. *Propagation in website pair*

Tables 7–9 present quantitatively the results of comparing webpages for propagation in (1:1) country-specific website pair in content categories "Corporate

Table 7. Propagation for content categories "Corporate Information" in website pair.

|    | IN | AU | UK | IE | US | CA | ME | ZA | Propagation | | | | Comparison of webpage |
|----|----|----|----|----|----|----|----|----|-----|-----|-----|-----|-----|
|    |    |    |    |    |    |    |    |    | Yes | %   | No  | %   |     |
| IN | —  | 15 | 15 | 14 | 12 | 13 | 14 | 15 | 98  | 70  | 42  | 30  | 140 |
| AU | 15 | —  | 15 | 15 | 13 | 14 | 14 | 16 | 87  | 73  | 33  | 27  | 120 |
| UK | 15 | 15 | —  | 17 | 12 | 15 | 15 | 16 | 75  | 75  | 25  | 25  | 100 |
| IE | 14 | 15 | 17 | —  | 12 | 15 | 15 | 16 | 58  | 73  | 22  | 27  | 80  |
| US | 12 | 13 | 12 | 12 | —  | 11 | 11 | 13 | 35  | 58  | 25  | 42  | 60  |
| CA | 13 | 14 | 15 | 15 | 11 | —  | 14 | 15 | 29  | 73  | 11  | 27  | 40  |
| ME | 14 | 14 | 15 | 15 | 11 | 14 | —  | 15 | 15  | 75  | 5   | 25  | 20  |
| ZA | 15 | 16 | 16 | 16 | 13 | 15 | 15 | —  | —   | —   | —   | —   | —   |
|    |    |    |    |    |    |    |    |    | **397** | **71** | **163** | **29** | **560** |

Table 8. Propagation for content categories "Product Information" in website pair.

|    | IN | AU | UK | IE | US | CA | ME | ZA | Propagation |    |     |    | Comparison of webpage |
|----|----|----|----|----|----|----|----|----|-----|----|-----|----|-----|
|    |    |    |    |    |    |    |    |    | Yes | %  | No  | %  |     |
| IN | —  | 8  | 9  | 8  | 5  | 6  | 8  | 8  | 52  | 37 | 88  | 63 | 140 |
| AU | 8  | —  | 9  | 8  | 4  | 6  | 7  | 8  | 42  | 35 | 78  | 65 | 120 |
| UK | 9  | 9  | —  | 11 | 5  | 6  | 8  | 8  | 38  | 38 | 62  | 62 | 100 |
| IE | 8  | 8  | 11 | —  | 6  | 6  | 7  | 8  | 27  | 34 | 53  | 66 | 80  |
| US | 5  | 4  | 5  | 6  | —  | 5  | 5  | 5  | 15  | 25 | 45  | 75 | 60  |
| CA | 6  | 6  | 6  | 6  | 5  | —  | 8  | 8  | 16  | 40 | 24  | 60 | 40  |
| ME | 8  | 7  | 8  | 7  | 5  | 8  | —  | 10 | 10  | 50 | 10  | 50 | 20  |
| ZA | 8  | 8  | 8  | 8  | 5  | 8  | 10 | —  | —   | —  | —   | —  | —   |
|    |    |    |    |    |    |    |    |    | 200 | 36 | 360 | 64 | 560 |

Table 9. Propagation for content categories "Customer Support Information" in website pair.

|    | IN | AU | UK | IE | US | CA | ME | ZA | Propagation |    |     |    | Comparison of webpage |
|----|----|----|----|----|----|----|----|----|-----|----|-----|----|-----|
|    |    |    |    |    |    |    |    |    | Yes | %  | No  | %  |     |
| IN | —  | 5  | 5  | 5  | 4  | 5  | 5  | 5  | 34  | 24 | 106 | 76 | 140 |
| AU | 5  | —  | 6  | 5  | 4  | 4  | 5  | 6  | 30  | 25 | 90  | 75 | 120 |
| UK | 5  | 6  | —  | 7  | 4  | 5  | 6  | 7  | 29  | 29 | 71  | 71 | 100 |
| IE | 5  | 5  | 7  | —  | 4  | 5  | 5  | 6  | 20  | 25 | 60  | 75 | 80  |
| US | 4  | 4  | 4  | 4  | —  | 4  | 4  | 4  | 12  | 20 | 48  | 80 | 60  |
| CA | 5  | 4  | 5  | 5  | 4  | —  | 5  | 5  | 10  | 25 | 30  | 75 | 40  |
| ME | 5  | 5  | 6  | 5  | 4  | 5  | —  | 7  | 7   | 35 | 13  | 65 | 20  |
| ZA | 5  | 6  | 7  | 6  | 4  | 5  | 7  | —  | —   | —  | —   | —  | —   |
|    |    |    |    |    |    |    |    |    | 142 | 25 | 418 | 75 | 560 |

Information", "Product Information" and "Customer Support Information" respectively. The coupling in sharing is represented from the occurrences of (a) propagation in website pair and (b) no propagation. The higher occurrences of propagation for specific content categories indicates high coupling between websites while sharing content for specific categories. Comparing the webpages, the differences in coupling in country-specific website pair with respect to categories are revealed.

As illustrated in Table 10, out of 560 comparisons of webpages in "Corporate Information" among 28 country-specific website pairs, 71% of cases with propagation occurring in website pairs are identified which suggests high coupling while sharing corporate related information. Contrary to this, 75% of cases with no propagation in website pairs are identified for "Customer Support Information" which suggests low coupling while sharing support related information. Similarly, for "Product Information" though the occurrences of no propagation are higher 64%, the differences with occurrences of propagation are comparable (only 28% while in other categories the difference is $> 50\%$). The coupling in a website pair while sharing product related information tends to be neutral.

Table 10. Summary on occurrences of propagation for content categories.

| Content category | Propagation in website pair | | No propagation in website pair | | Comparison of webpage |
|---|---|---|---|---|---|
| | N | % | N | % | |
| (a) Corporate information | 397 | 71 | 163 | 29 | 560 |
| (b) Product information | 200 | 36 | 360 | 64 | 560 |
| (c) Customer support information | 142 | 25 | 418 | 75 | 560 |
| Total | 739 | 44 | 941 | 56 | 1680 |

Figure 4(b) presents the comparative view on the occurrences of propagation for the content categories in website pairs. The coupling in websites for sharing content is higher for content that is globally suitable such as corporate related information and decreases as the scales of the content narrows down within a country. From examining the occurrences of propagation and no propagation in (1:1) country-specific website pair; the coupling in sharing for specific content categories is compiled in this study.

### 5.2.1. Coupling in sharing

Priorities for content consistency with respect to content categories while sharing are inferred from coupling between websites. High coupling in website pairs for sharing corporate related information suggests that content contribution occurs frequently and so consistency has to be strictly enforced while sharing such content. Low coupling in website pairs for sharing support related information suggests that content contribution occurs less frequently and so the policy for consistency is not strictly enforced while sharing such content. Similarly, neutral coupling in website pairs for sharing product related information suggests that the policy for consistency is to be moderately enforced while sharing such content.

Coupling in sharing is useful in setting priorities for content consistency in the presence of several content categories. As identified in this study, the content consistency for corporate related information has higher priorities in comparison to product related and customer support related information during information sharing.

From propagation in (1:7) websites and (1:1) website pairs, the results in Table 11 compile the traits in sharing content from specific categories: scales and coupling which are useful to promote consistent information sharing. In the next section, the hypothesis is verified.

Table 11. Summary on scales and coupling in sharing for content categories.

| Content category | Scales in sharing | Coupling in sharing |
|---|---|---|
| (a) Corporate information | Global | High |
| (b) Product information | Local and regional | Neutral |
| (c) Customer support information | Local | Low |

## 6. Hypothesis Verification

As illustrated in Tables 6 and 10, the differences in propagation both in terms of scale and number of occurrences among country-specific websites are identified for the content categories. Suitability for sharing corporate related information globally in all countries, sharing customer support related information locally within specific country and sharing product related information both regionally and locally suggest the scales in sharing that restrict the publication of content and their description in specific languages. Similarly, the occurrences of propagation in websites are also found to vary with respect to content categories. High coupling between websites while sharing corporate related information indicates that they are shared more frequently and so consistency has to be strictly enforced while sharing such content.

From identifying traits such as scales and coupling in sharing that vary with content categories, this study verified propagation among county-specific websites constrained for specific categories during information sharing. Integrating scales to achieve a mechanism for consistent information sharing for propagating content updates is proposed in the next section.

## 7. Mechanism for Consistent Information Sharing Among Countries

This section details the proposal of mechanism for consistent information sharing by enabling propagation of up-to-date content among communities conforming to global, regional and local communities. The focus is also on a mechanism that is simple and applicable in both monolingual and multilingual cases.

### 7.1. *Proposed pattern of sharing*

From previous study (Pariyar *et al.*, 2015), the cases of propagation in (1:7) country-specific websites: (a) propagation to all country-specific websites, (b) some websites and (c) propagation to none, are generalised into patterns of sharing as below.

(a) **Internationalisation:** Previous studies (Esselink, 2000) have presented views on internationalisation as the process of generalising a product for handling multiple languages and cultural conventions. With respect to content shared among country-specific websites, this suggests content suitable at a global scale. This also indicates on the propagation of content among entire countries from the publication of content globally and description in all available languages offered in global organisations. Figure 5(a) resembles the internationalisation pattern of sharing from propagation to countries in all geographical regions. By integrating the internationalisation pattern, the content updates can be propagated globally to ensure consistency when sharing corporate related information among countries in their official languages.

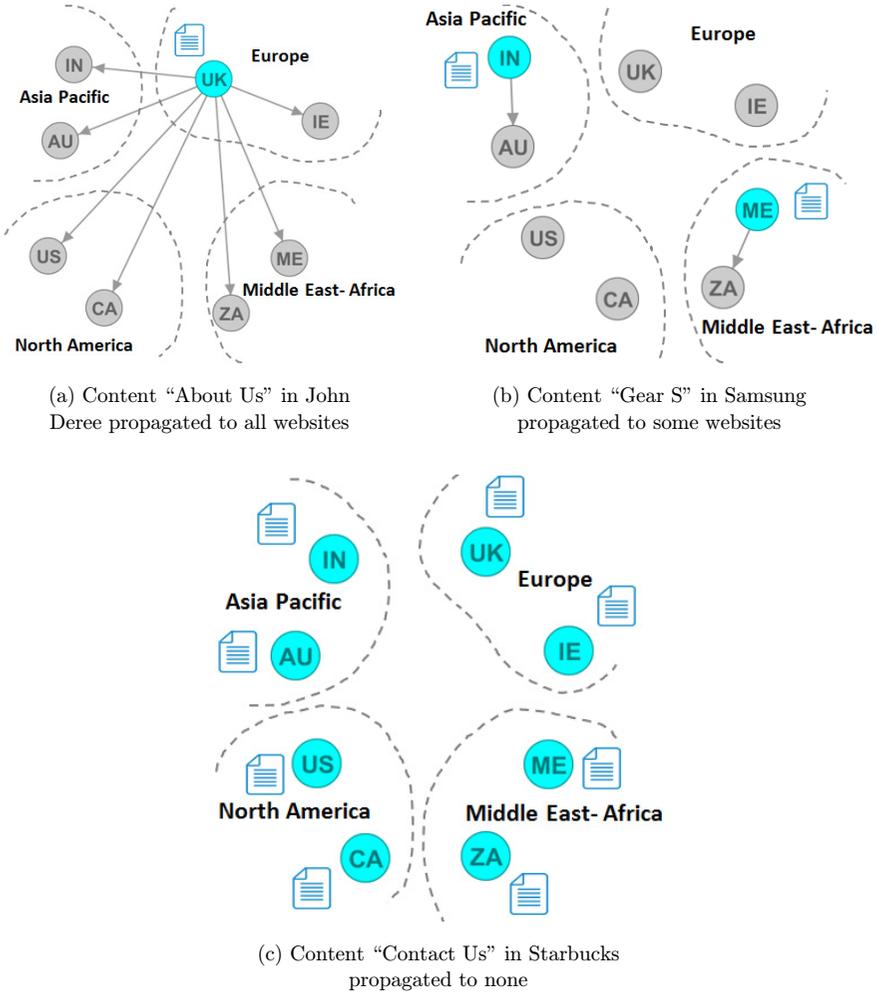

(a) Content "About Us" in John Deree propagated to all websites

(b) Content "Gear S" in Samsung propagated to some websites

(c) Content "Contact Us" in Starbucks propagated to none

Fig. 5. Propagation among country-specific websites in global brands.

(b) **Regionalisation:** In context to globalisation, the view on regionalisation represents a world that becomes less interconnected with a stronger regional focus and is interesting for researches on market segmentation (Rugman and Verbeke, 2004). Inter-regional and intra-regional suitability in content shared as identified for "Product Information" generalises as regionalisation pattern in the information sharing regionally among countries (Fig. 5(b)).

(c) **Localization:** The view on localisation is towards the process of making a product linguistically and culturally appropriate to the target locale (country/region and language) where it will be used and sold (Esselink, 2000). For content shared among country-specific websites, localisation suggests on the suitability of content at local scale without any necessity for the propagation of content updates among country-specific websites (Fig. 5(c)). However, integrating localisation in information sharing ensures the propagation of content updates in languages locally within a country. Local consistency when sharing Customer Support Information is achieved by integrating localisation pattern during information sharing.

The combination of proposed pattern of sharing is also possible to achieve with consistent information sharing when sharing content with components, each globally, locally and regionally suitable. For example, glocalisation due to content communicating both globally and locally (Maynard and Tian, 2004) is achieved by integrating proposed internationalisation and localisation patterns so that content updates are propagated globally and locally where needed. The combination of such pattern also allows globally relevant information to be reused among countries while locally relevant information from one country serves as reference to another country. The global and local dimension of knowledge development and sharing as described in Adenfelt and Lagerström (2006) is also illustrated from such pattern.

Integrating pattern of sharing (a) internationalisation, (b) regionalisation, (c) localisation and their combinations in propagating content updates restricted to global, regional or local communities and their specific languages, consistency during information sharing is achieved. Such pattern of sharing can be applied to content either (i) automatically by employing text mining approaches to identify specific categories or (ii) manually by generating policy while sharing among countries. The proposed pattern of sharing integrated with inconsistency detection mechanism in our previous study (Pariyar *et al.*, 2014) offers inconsistency management for multi-language knowledge sharing system. Multilingual service platform such as language grid (Ishida, 2011) which offers language resources such as dictionaries, machine translations is useful as components in such systems. The formalism of proposed pattern of sharing along with an example to illustrate the pattern applied during information sharing is presented next.

### 7.2. *Formalising rules*

To describe the rules associated with pattern of sharing, we will use the following notations.

Collection of website W is published in a global organisation where each country-specific website $W_j \in$ W is targeted for specific country $j$. Collection of language L is used in the organisation with an official language $L_j \in$ L for a specific country. $W_j^i$ represents country-specific website for country $j$ offering content in language $i$. R represents a geographic region to which a country $j$ belongs.

Following this formalisation, a country-specific website for Canada is $W_{j=ca}$ and the official languages of Canada $L_{j=ca} = \{en, fr\}$ are English and French.

The constructs used in formalising rules are as following. Content shared among countries with a specific pattern is represented with *isSharedApplying(Content, Pattern)*.

Table 12. Rules in pattern of sharing.

| Pattern | Rule in collaboration |
| --- | --- |
| Internationalisation | **Rule 1:** $\forall x : isContent(x) \land isSharedApplying\,(x, Internationalisation) \Rightarrow isPublishedIn(x, W) \land isDescribedIn(x, L)$ <br> **Description:** Content shared applying internationalisation pattern is published in entire websites and is described in entire languages offered among the websites. |
| Localisation | **Rule 2:** $\forall x, \exists j : isContent(x) \land isSharedApplying\,(x, Localisation) \Rightarrow isPublishedIn(x, W_j) \land isDescribedIn(x, L_j)$ <br> **Description:** Content shared applying localisation pattern is published in the website of specific country $j$ and described only in the languages offered in that particular country. |
| Regionalisation | **Rule 3:** $\forall x, \exists j \in R : isContent(x) \land isSharedApplying\,(x, Regionalisation) \Rightarrow isPublishedIn(x, W_j) \land isDescribedIn(x, L_j)$ <br> **Description:** Content shared applying regionalisation pattern is publish at the specific website of country $j$ belonging to specific region R and described in the languages offered in countries from the region R. |

Content published in a website and described in specific language is represented with $isPublishedIn(Content, Website)$ and $isDescribedIn(Content, Language)$ respectively.

Rules associated with the pattern of sharing are illustrated in Table 12. The publication of content at specific websites and their description in specific languages represent the scales that restrict the propagation of content updates. The pattern of sharing applied in a collaborative setting is explained next.

### 7.3. *Applying pattern of sharing*

For illustration purposes, web mangers managing websites for Canada (CA), United Kingdom (UK), United States (US), India (IN), Nepal (NP) with content offered in their official languages $L_{j=ca} = \{en, fr\}$, $L_{j=uk} = \{en\}$, $L_{j=us} = \{en\}$, $L_{j=in} = \{hi, en\}$ and $L_{j=np} = \{np\}$ are considered as participants in information sharing (Fig. 6). In this example, "English" language is shared among UK, US and India; while both India and Canada also have multiple official languages. Consistent information sharing among countries is required in both shared and unshared languages where needed. Figure 6(a) represents internationalisation pattern applied for sharing by the web manager for United States while sharing content with remaining countries. From Rule 1 in Table 12, the content is published in the United States and the description of content is shared in several languages (English, French, Hindi, Nepali) offered in websites for UK, CA, IN and NP. Updating content at any country-specific websites for United Kingdom, Canada, India, Nepal, United States and in any languages will result in propagating updates to at all websites from such pattern. This is depicted from the undirected propagation among the countries.

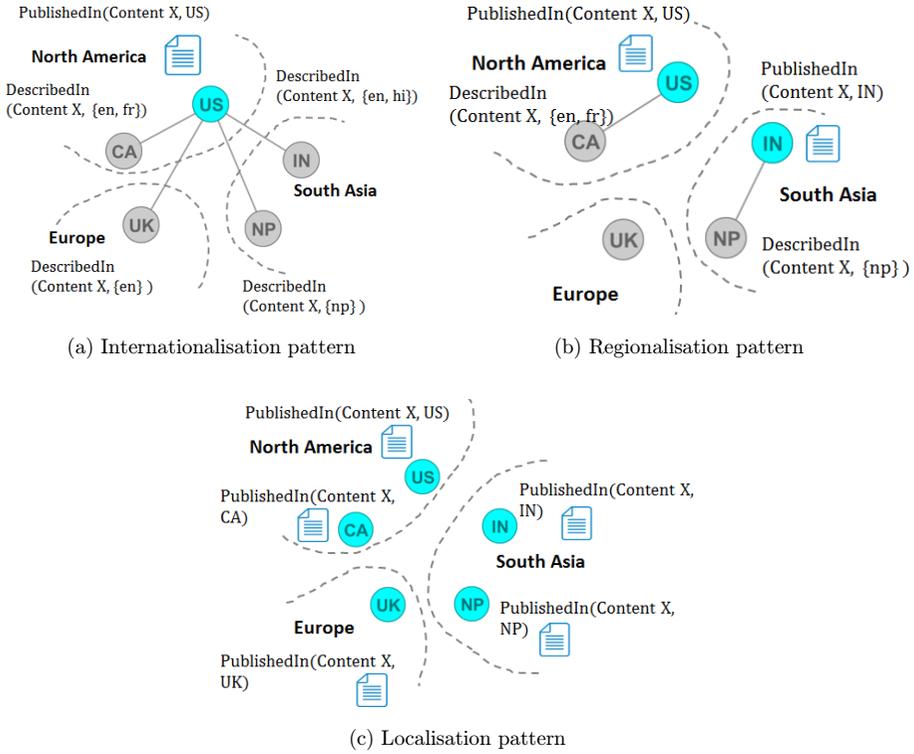

Fig. 6. Pattern of sharing applied in information sharing.

Synchronisation of content updates among countries in several official languages (both shared and unshared) is achieved from restricting the propagation globally.

Information sharing in Fig. 6(b) represents regionalisation pattern applied by web managers for the United States and India to share content updates regionally with the Canada and Nepal respectively. From Rule 2 in Table 12, the description of content is shared in several languages (English and French) offered among US and Canada and in languages (English, Hindi and Nepali) offered among Nepal and India. Content updates are shared regionally between countries such as within South Asia rather than with other regions. Inconsistency in regionally shared content is avoided from the propagation of content updates limited to countries within regions and languages offered in those regions.

Figure 6(c) represents localisation pattern applied for information sharing locally. Such pattern is useful for restricting the propagation of content updates locally for countries offering multiple official languages. For example, content updates are synchronised locally in languages (English and Hindi) for India without sharing with the remaining countries. With such a pattern, the content is not shared with other countries even when a language is shared and updates are confined within a country and only in its languages. Inconsistency in local languages within a country is avoided by applying localisation pattern during information sharing.

From the proposed pattern of sharing we illustrated, the propagation of content updates confined to specific communities globally, regionally and locally to achieve consistent information sharing. Examining propagation among country-specific websites, we identified traits in sharing such as scales and coupling in sharing that differ with respect to content categories. Integrating pattern of sharing to specific content categories either automatically or manually, inconsistency from missing content, outdated content and conflicting content among countries which are considered as problematic in information sharing are avoided by propagation before even advancing for translation.

## 8. Discussion and Future Work

From the verification of hypothesis, traits conforming to scales and coupling in sharing are identified for specific content categories. Global scales in sharing corporate related information, local scales while sharing customer support related information and both regional and local scales suitable for sharing product related information illustrated that propagation among countries varies depending on content belonging to specific categories. Similarly, the number of occurrences of propagation is also found to vary with respect to content categories. Corporate related information tends to be propagated more frequently in comparison to other categories which suggested high coupling between websites while sharing such content. Coupling in sharing is also useful in setting priorities for content consistency in the presence of several content categories. By incorporating content scales with pattern of sharing (a) internationalisation, (b) regionalisation, (c) localisation and their combination for consistent information sharing, this study proposed a simple mechanism applicable in both monolingual and multilingual cases. Consistency is achieved by propagating content updates restricted to global, regional or local communities and their specific languages. The contributions from this research are summarised next.

Though past studies are known to employ techniques for multilingual content synchronisation, this study contributed in the earlier phases of information sharing when there is a need to restrict sharing among specific communities and their languages. Also, in past studies, it was not known how different categories of content had to be synchronised. By analysing content from different categories, this study contributed by identifying the consistency needs for specific content which is useful for apply existing techniques in the later phases of information sharing.

Past studies are also known to target information disparities mostly in multilingual Wikipedia articles where content is shared in diverse topics. Inside an organisation, content is limited in scope within its business domain and targeted at specific communities such as investors, business partners and customers; therefore, approaches in past studies cannot be directly applied to resolve inconsistency. This study contributed by expanding our understanding of sharing in a business domain so as to prepare consistency policy most suited for organisations. Further, past studies examined content and design features in websites that represented headquarter offices of global brands and the results were generalised over cultural groups or industries. This study contributed to existing studies by examining country-specific websites inside each global brand which opened our understating of sharing among headquarter offices and their branch offices. However, there are some limitations which are summarised next.

The data samples for this study included 10 global brands, 80 websites, 480 comparisons in complete graph and 1680 comparisons in website pairs. Though the samples were promising to explain traits in sharing, the limitations are due to the sampling of websites. For this study, country-specific websites that offered content only in English language were examined for sharing which ignored the pattern of sharing that could have explained cross language sharing. The reason for this is unfamiliarity with other languages. However, extensive analysis involving language experts to examine content in different languages can be a promising future direction.

The sample of country-specific websites chosen to represent each geographic region is also limited which may not fully explain the sharing phenomena involved in that region. By increasing the number of country-specific websites in each region, the result would be convincing to explain the coupling among websites inside that region. Similarly, the sample of content categories is limited to corporate related, customer support related and product related information which can be further expanded to include information such as corporate social responsibility, blogs, career information and so. This study also qualitatively compared content in webpages between country-specific websites to check whether they share same and paraphrased content. The techniques that involve lexical mining by extracting geographic references as in (Quercini *et al.*, 2010) offer a direction towards examining geographically relevant content between countries.

Future work will focus on integrating the proposed pattern of sharing with inconsistency detection mechanism in Pariyar *et al.* (2014) to design a multi-language knowledge sharing system. In doing so, we will also evaluate the practicability of the system in disseminating information to global, regional and local communities.

## 9. Conclusion

This research qualitatively studied information sharing among countries as depicted by content shared in their country-specific websites. From examining content published in country-specific websites inside global brands, inconsistency in both shared and unshared languages is illustrated. Further content propagation among websites is compared in a complete graph and website pairs to determine traits in sharing among countries. Examining propagation in (1:7) country-specific websites revealed scales in sharing that varied with specific content categories. Results suggested that corporate related information tends to be shared globally and customer support related information tends to be shared locally while product related information tends to be locally and regionally suitable for sharing. Examining propagation in (1:1) country-specific website pairs revealed coupling in sharing due to differences in the occurrences of propagation for specific content categories. Results showed tendency for high coupling in websites while sharing corporate related information and suggested content consistency to be strictly enforced for such content.

Towards promoting consistency, this study also proposed a mechanism to restrict propagation of content updates with pattern of sharing (a) internationalisation, (b) regionalisation (c) localisation and their combinations. Such technique is useful in the earlier phases of information sharing as it allows propagating content updates where necessary before existing techniques becomes applicable. Compared to past studies that focused primarily on Wikipedia articles with content in diverse topics, this study contributed by expanding our understanding on the current state of information sharing in a business domain.

### Acknowledgement

This research was partially supported by a Grant-in-Aid for Scientific Research (S) (24220002) from Japan Society for the Promotion of Science.

### References

Adar, E, M Skinner and DS Weld (2009). Information arbitrage across multi-lingual Wikipedia. In *Proceedings of the Second ACM International Conference on Web Search and Data Mining*, pp. 94–103. New York: ACM.

Adenfelt, M and K Lagerström (2006). Knowledge development and sharing in multinational corporations: The case of a centre of excellence and a transnational team. *International Business Review*, 15(4), 381–400.

Almeida, P and A Phene (2004). Subsidiaries and knowledge creation: The influence of the MNC and host country on innovation. *Strategic Management Journal*, 25(8–9), 847–864.

Bey, Y, C Boitet and K Kageura (2006). The TRANSBey prototype: An online collaborative wiki-based cat environment for volunteer translators. In *LREC-2006: Fifth International Conference on Language Resources and Evaluation*, Genoa, Italy, pp. 49–54.

Blenkinsopp, J and MS Pajouh (2010). Lost in translation? Culture, language and the role of the translator in international business. *Critical perspectives on International Business*, 6(1), 38–52.

Bronner, A, M Negri, Y Mehdad, A Fahrni and C Monz (2012). Cosyne: Synchronizing multilingual wiki content. In *Proceedings of the Eighth Annual International Symposium on Wikis and Open Collaboration*, pp. 1–33. New York: ACM.

Cabrera, A, WC Collins and JF Salgado (2006). Determinants of individual engagement in knowledge sharing. *The International Journal of Human Resource Management*, 17(2), 245–264.

Charles, M and R Marschan-Piekkari (2002). Language training for enhanced horizontal communication: A challenge for MNCs. *Business Communication Quarterly*, 65(2), 9–29.

Cormican, K and L Dooley (2007). Knowledge sharing in a collaborative networked environment. *Journal of Information & Knowledge Management*, 6(2), 105–114.

Daryanto, A, H Khan, H Matlay and R Chakrabarti (2013). Adoption of country-specific business websites: The case of UK small businesses entering the Chinese market. *Journal of Small Business and Enterprise Development*, 20(3), 650–660.

Esselink, B (2000). *A Practical Guide to Localization*, Vol. 4. Amsterdam: John Benjamins Publishing.

Feely, AJ and AW Harzing (2003). Language management in multinational companies. *Cross Cultural Management: An International Journal*, 10(2), 37–52.

Fredriksson, R, W Barner-Rasmussen and R Piekkari (2006). The multinational corporation as a multilingual organization: The notion of a common corporate language. *Corporate Communications: An International Journal*, 11(4), 406–423.

Frické, M (2009). The knowledge pyramid: A critique of the DIKW hierarchy. *Journal of Information Science*, 35(2), 131–142.

Gupta, A, E Mattarelli, S Seshasai and J Broschak (2009). Use of collaborative technologies and knowledge sharing in co-located and distributed teams: Towards the 24-h knowledge factory. *The Journal of Strategic Information Systems*, 18(3), 147–161.

Hall, S (1997). The local and the global: Globalization and ethnicity. *Cultural Politics*, 11, 173–187.

He, W and KK Wei (2009). What drives continued knowledge sharing? An investigation of knowledge-contribution and-seeking beliefs. *Decision Support Systems*, 46(4), 826–838.

Hofstede, G (1984). *Culture's Consequences: International Differences in Work-Related Values*, Vol. 5. Beverly Hills, CA: Sage Publication.

Huizingh, EK (2000). The content and design of web sites: An empirical study. *Information & Management*, 37(3), 123–134.

Inaba, R, Y Murakami, A Nadamoto and T Ishida (2007). Multilingual communication support using the language grid. In *Intercultural Collaboration*, pp. 118–132, Berlin: Springer.

Ishida, T (2011). *The Language Grid.* Berlin: Springer.

Kale, SH (1991). Culture-specific marketing communications: An analytical approach. *International Marketing Review*, 8(2), 18–30.

Kumaran, A, N Datha, B Ashok, K Saravanan, A Ande, A Sharma, S Vedantham, V Natampally, V Dendi and S Maurice (2010). WikiBABEL: A system for multilingual wikipedia content. *In Proceedings of the collaborative Translation Workshop: Technology, Crowdsourcing, and the Translator Perspective at the AMTA 2010 Conference, Denver, Colorado.*

Lagerström, K and M Andersson (2003). Creating and sharing knowledge within a transnational team-the development of a global business system. *Journal of World Business*, 38(2), 84–95.

Liang, TP, CC Liu and CH Wu (2008). Can social exchange theory explain individual knowledge-sharing behavior? A meta-analysis. In *ICIS 2008 Proceedings*, paper 171.

Lionbridge (2009). Building a global web strategy best practices for developing your international online brand. Available at http://www.slideshare.net/Lionbridge. Accessed on 5 March 2015.

Main, L (2001). The global information infrastructure: Empowerment or imperialism? *Third World Quarterly*, 22(1), 83–97.

Markus, ML (2001). Toward a theory of knowledge reuse: Types of knowledge reuse situations and factors in reuse success. *Journal of Management Information Systems*, 18(1), 57–94.

Marschan-Piekkari, R, D Welch and L Welch (1999). Adopting a common corporate language: IHRM implications. *International Journal of Human Resource Management*, 10(3), 377–390.

Maynard, M and Y Tian (2004). Between global and glocal: Content analysis of the Chinese web sites of the 100 top global brands. *Public Relations Review*, 30(3), 285–291.

Monz, C, V Nastase, M Negri, A Fahrni, Y Mehdad and M Strube (2011). Cosyne: A framework for multilingual content synchronization of wikis. In *Proceedings of the 7th International Symposium on Wikis and Open Collaboration*, Mountain View, CA, USA, pp. 217–218, 3–5 October.

Nohria, N and S Ghoshal (1994). Differentiated fit and shared values: Alternatives for managing headquarters-subsidiary relations. *Strategic Management Journal*, 15(6), 491–502.

Pariyar, A, Y Murakami, D Lin and T Ishida (2014). Inconsistency detection in multilingual knowledge sharing. *Journal of Information & Knowledge Management*, 13(4), 1450033.

Pariyar, A, Y Murakami, D Lin and T Ishida (2015). Content sharing in global organization: A cross-country Perspective. In *2014 ASE BigData/SocialInformatics Conference*, Harvard University, 14–16 December.

Quercini, G, H Samet, J Sankaranarayanan and MD Lieberman (2010). Determining the spatial reader scopes of news sources using local lexicons. In *Procedings of the 18th SIGSPATIAL International Conference on Advances in Geographic Information Systems*, pp. 43–52. New York: ACM.

Robbins, SS and AC Stylianou (2003). Global corporate web sites: An empirical investigation of content and design. *Information & Management*, 40(3), 205–212.

Rowley, JE and RJ Hartley (eds.) (2008). *Organizing Knowledge: An Introduction to Managing Access to Information.* Aldershot: Ashgate Publishing.

Rugman, AM and A Verbeke (2004). A perspective on regional and global strategies of multinational enterprises. *Journal of International Business Studies*, 35(1), 3–18.

Ryan, SD, JC Windsor, B Ibragimova and VR Prybutok (2010). Organizational practices that foster knowledge sharing: Validation across distinct national cultures. *Informing Science: The International Journal of an Emerging Transdiscipline*, 13, 139–164.

Singh, N, V Kumar and D Baack (2005). Adaptation of cultural content: Evidence from B2C e-commerce firms. *European Journal of Marketing*, 39(1/2), 71–86.

Singh, N and A Pereira (2005). *The Culturally Customized Web Site*. New York: Routledge.

Singh, N, DR Toy and LK Wright (2009). A diagnostic framework for measuring Web-site localization. *Thunderbird International Business Review*, 51(3), 281–295.

Steinheider, B and S Al-Hawamdeh (2004). Team coordination, communication and knowledge sharing in SMEs and large organisations. *Journal of Information & Knowledge Management*, 3(3), 223–232.

Tanaka, R, Y Murakami and T Ishida (2011). Cascading translation services. In *The Language Grid*, pp. 103–117. Berlin: Springer.

Tixier, M (2005). Globalization and localization of contents: Evolution of major internet sites across sectors of industry. *Thunderbird International Business Review*, 47(1), 15–48.

Wagner, C (2004). Wiki: A technology for conversational knowledge management and group collaboration. *The Communications of the Association for Information Systems*, 13(1), 58.

Wallace, DP (2007). *Knowledge Management: Historical and Cross-Disciplinary Themes.* West Port, CT: Libraries Unlimited.

Welch, DE, LS Welch and R Marschan-Piekkari (2001). The persistent impact of language on global operations. *Prometheus*, 19(3), 193–209.

Yamashita, N, R Inaba, H Kuzuoka and T Ishida (2009). Difficulties in establishing common ground in multiparty groups using machine translation. In *Proceedings of the SIGCHI Conference on Human Factors in Computing Systems*, pp. 679–688. New York: ACM.

Yamashita, N and T Ishida (2006). Effects of machine translation on collaborative work. In *Proceedings of the 2006 20th Anniversary Conference on Computer Supported Cooperative Work*, pp. 515–524. New York: ACM.

Yunker, J (2002). *Beyond Borders: Web Globalization Strategies.* Berkeley, CA: New Riders.

Yunker, J (2014). The 2014 web globalization report card. Available at http://bytelevel.com/reportcard2014/. Accessed on 5 March 2015.

Zhao, H and Y Luo (2005). Antecedents of knowledge sharing with peer subsidiaries in other countries: A perspective from subsidiary managers in a foreign emerging market. *MIR: Management International Review*, 45(1), 71–97.